# AI-Generated Annotations Dataset for Diverse Cancer Radiology Collections in NCI Image Data Commons

## Authors
Gowtham Krishnan Murugesan[1], Diana McCrumb[1], Mariam Aboian[2], Tej Verma[2], Rahul Soni[1], Fatima Memon[2], Keyvan Farahani[3], Linmin Pei[4], Ulrike Wagner[4], Andrey Y. Fedorov[5], David Clunie[6], Stephen Moore[1], Jeff Van Oss[1]

**Affiliations**

1. BAMF Health, Grand Rapids, MI, USA
2. Yale School of Medicine, New Haven, CT, USA
3. National Institute of Health, Bethesda, MD, USA
4. Frederick National Laboratory for Cancer Research, Frederick, MD, USA
5. Brigham and Women's Hospital and Harvard Medical School, Boston, MA, USA
6. PixelMed Publishing, Bangor, PA, USA

corresponding author(s): Gowtham Krishnan Murugesan (gowtham.murugesan@bamfhealth.com)

## Abstract
The National Cancer Institute (NCI) Image Data Commons (IDC) offers publicly available cancer radiology collections for cloud computing, crucial for developing advanced imaging tools and algorithms. Despite their potential, these collections are minimally annotated; only 4% of DICOM studies in collections considered in the project had existing segmentation annotations. This project increases the quantity of segmentations in various IDC collections. We produced high-quality, AI-generated imaging annotations dataset of tissues, organs, and/or cancers for 11 distinct IDC image collections. These collections contain images from a variety of modalities, including computed tomography (CT), magnetic resonance imaging (MRI), and positron emission tomography (PET). The collections cover various body parts, such as the chest, breast, kidneys, prostate, and liver. A portion of the AI annotations were reviewed and corrected by a radiologist to assess the performance of the AI models. Both the AI's and the radiologist's annotations were encoded in conformance to the Digital Imaging and Communications in Medicine (DICOM) standard, allowing for seamless integration into the IDC collections as third-party analysis collections. All the models, images and annotations are publicly accessible.

## Background & Summary
The National Cancer Institute (NCI) Imaging Data Commons (IDC) is a comprehensive repository of publicly available cancer imaging data, image-derived insights, and associated resources[1]. These collections, drawn from diverse public imaging initiatives such as The Cancer Imaging Archive (TCIA)[2], are encoded in a standard representation in the IDC according to the Digital Imaging and Communications in Medicine (DICOM) standard. As of December 2023, IDC contains over 45 TB of publicly available cancer imaging data. Along with the data, IDC provides cloud-based tools to enable search, visualization, analysis and sharing of analysis results, which supports a continuous cycle of data nurturing, refinement, and enrichment, which is a key characteristic of data commons.

The Artificial Intelligence (AI) in Medical Imaging (AIMI) initiative strives to address the ever-persistent challenge of publicly available imaging data: the lack of comprehensive and high-quality annotations. The goal of this project, in the context of the AIMI initiative, was to curate AI-generated annotations to produce derived datasets from 11 distinct image collections in the IDC. The selected subset of tasks was chosen to ensure a diverse representation of various cancer types and imaging modalities, with a minimum of 2,000 images for annotation. The image collections encompass three modalities (CT, PET, and MR), and various annotation types, as detailed in Table 1. FDG PET/CT collections taken from The Cancer Genome Atlas (TCGA) program for Lung Adenocarcinoma (LUAD)[3] and Lung Squamous Cell Carcinoma (LUSC)[4], the Lung Cancer Diagnosis (LUNG-PET-CT-Dx)[5], Anti-PD-1 Immunotherapy Lung[6], the Reference Image Database to Evaluate Therapy Response PET/CT subgroup (RIDER Lung PET-CT)[7], and the Non-Small Cell Lung Cancer (NSCLC) for Radiogenomics[8–10] and the American College of Radiology Imaging Network non-small cell lung cancer FDG PET (ACRIN-NSCLC-FDG-PET)[11,12] were used for both lung and lung tumor annotations. The CTs of the previously listed lung collections, excluding ACRIN- NSCLC, were also used for the annotation of lung nodules. Breast tumor annotations were generated on the FDG PET/CT breast dataset from the Quantitative Imaging Network (QIN)[13,14]. CT annotations for kidneys, kidney tumors, and kidney cysts were generated for the Kidney Renal Clear Cell Carcinoma Collection (KIRC)[15] from the TCGA. MRI prostate-only annotations were generated for the ProstateX[16,17] collections. Liver-only annotations were generated for both CT and MRI data from the Liver Hepatocellular Carcinoma (LIHC)[18] collection of the TCGA.

*Table 1: IDC collections used for AI-assisted annotations.*

| Collection | No. of Cases | Modality of interest (MOI) | No. of Cases w/ MOI | Annotation Task (associated modality) |
|---|---|---|---|---|
| **TCGA-LUAD**[3] | 560 | CT, PET | 68 | Lungs (CT), tumors (PET/CT), and nodules (CT) |
| **TCGA-LUSC**[4] | 504 | CT, PET | 37 | Lungs (CT), tumors (PET/CT), and nodules (CT) |
| **LUNG-PET-CT-Dx**[5] | 355 | CT, PET | 355 | Lungs (CT), tumors (PET/CT), and nodules (CT) |
| **Anti-PD-1-Lung**[6] | 242 | CT, PET | 242 | Lungs (CT), tumors (PET/CT), and nodules (CT) |
| **RIDER Lung PET-CT**[7] | 243 | CT, PET | 243 | Lungs (CT), tumors (PET/CT), and nodules (CT) |
| **NSCLC-Radiogenomics**[8–10] | 211 | CT, PET | 211 | Lungs (CT), tumors (PET/CT), and nodules (CT) |
| **ACRIN-NSCLC-FDG-PET**[11,12] | 46 | CT, PET | 46 | Lungs (CT), and tumors (PET/CT) |
| **QIN-Breast**[13,14] | 68 | CT, PET | 43 | Tumors (PET/CT) |
| **TCGA-KIRC**[15] | 537 | CT | 237 | Kidneys (CT), tumors CT), and cysts (CT), |
| **ProstateX**[16,17] | 346 | MRI | 346 | Prostate (MRI) |
| **TCGA-LIHC**[18] | 377 | CT, MRI | 97 | Liver (CT), Liver (MRI) |

Some collections primarily contain histopathological data rather than radiological imaging data[3,4,18]. The number of cases available for each collection for AI-generated annotations was limited to the number of cases that contained the imaging region of the annotation task of interest within the modality of interest. Figure 1 shows the modality breakdown of the total number of imaging studies within the total number of suitable cases for each collection.

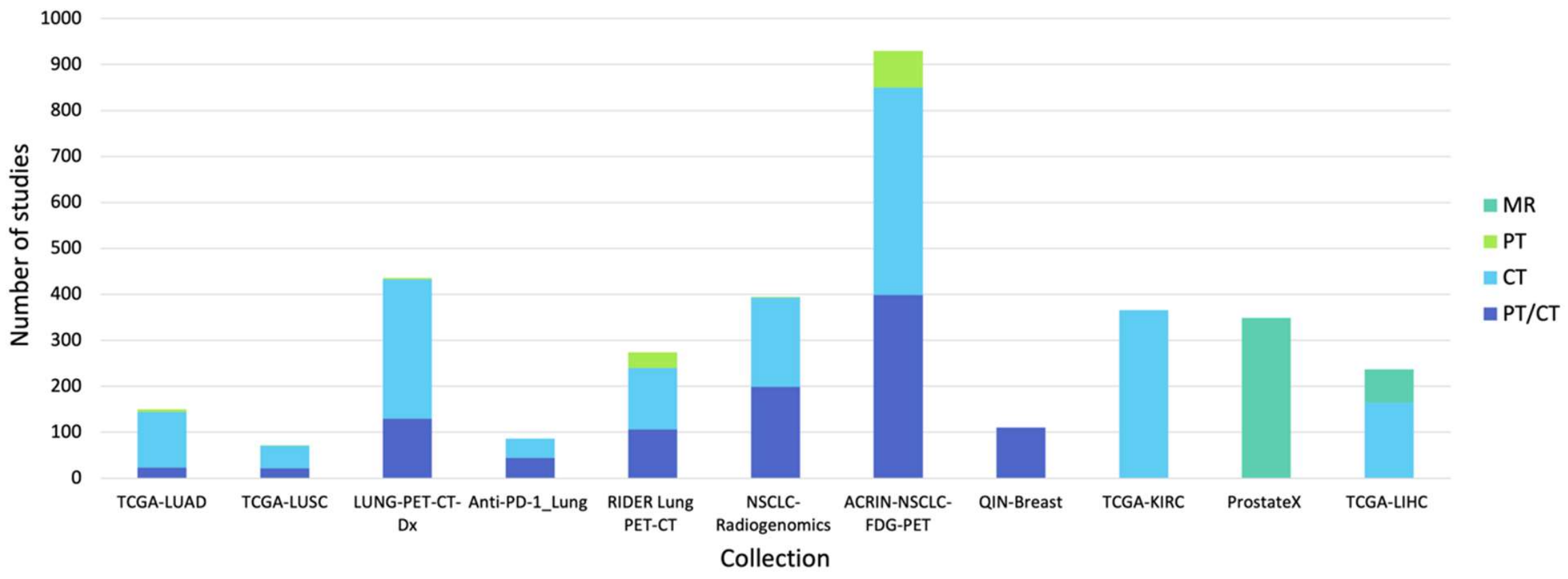

*Figure 1: Modality breakdown for the total number of imaging studies within the number of suitable cases for each collection.*

Another component of this project was to ensure the relationship between the annotations and their represented image dataset was maintained and queryable. To achieve this, all AI-generated annotations were encoded in conformance with the DICOM standard[19]. This makes them easily accessible within the existing tools and workflows of the IDC. The annotations are also accompanied by cloud-ready analysis workflows, embodied within Google Collaboratory notebooks. These resources empower users not only to recreate the dataset but also offer practical guidance on querying, visualizing, and transforming DICOM-encoded objects into alternative representations.
In summary, this project underscores the pivotal role played by the IDC platform in ameliorating annotation deficiencies within publicly available cancer imaging datasets. Through the introduction of AI-generated annotations, we have not only enriched the utility of these collections but also exemplified the collaborative potential of AI in advancing the frontiers of cancer imaging research.

## Methods

This section is structured according to the annotation type listed in Table 1 and follows the order displayed in Figure 2. Each task has its own uniquely catered methodology but follows the general workflow outline in Figure 2A.

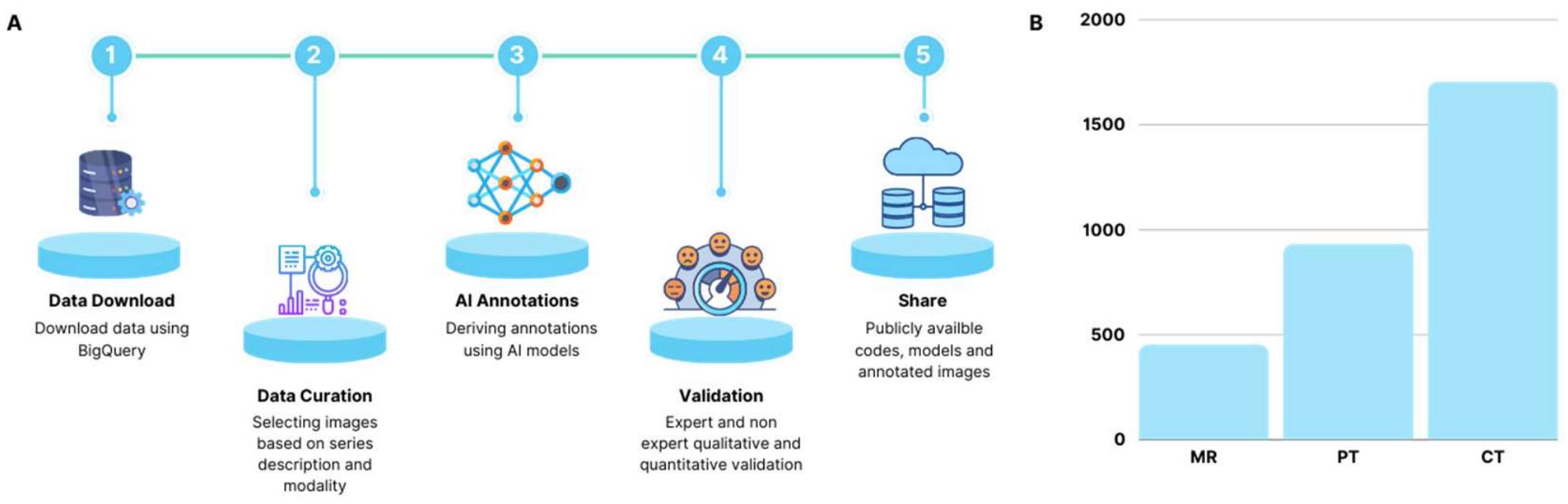

*Figure 2: Overview of AI-Generated Annotations: A). Illustration of AI-generated annotation tasks and their associated modalities, B). Workflow for the annotation tasks, C). Distribution of the total number of DICOM series per modality of interest across 11 collections*

All collections publicly available in Imaging Data Commons and were downloaded from Google Cloud Platform using BigQuery queries. Through data curation, a total of 3091 radiologic DICOM series (934 PET, 1704 CT, and 453 MRI) across all 11 collections (Figure 2B) were identified for annotations. All datasets in the experiment are deidentified.
Supervised deep learning AI models, ensembled from five-fold cross validation models using the nnU-Net[20] framework, were trained from a combination of the imaging data available in the IDC and additional publicly available collections. Following our previous observation[21] that multi-task nnU-Net[20] models outperformed single-label models in detecting whole body FDG PET/CT lesions, the nnU-Net[20] models for the PET/CT annotations were trained as multi-task models that detects multiple organs in addition to lesions. The multi-task labels for the training data were assembled from publicly available expert annotations, as well as output predictions generated by a different model, TotalSegmentator[22]. The training sets differ between models and a description of each training set is provided in subsequent sections.

To evaluate the quality and accuracy of the AI predictions, approximately 10% of the data was reviewed and corrected for quantifiable quality control by both a board-certified radiologist (expert) and an annotation specialist (non-expert). The annotation specialist has medical knowledge and a passing familiarity with radiology scans but is not a certified expert. The reviewers were provided with the cases and ai segmentation files in NIFTI format. The reviewers used their preferred viewer (ITKSnap[1]) to load the case and ai segmentation. Both the radiologist and the annotation specialist rated the AI predictions per case on a Likert Scale to assess their quality, as described in Table 2. Likert score ratings were assigned to each case, evaluating the overall quality of annotations across all labels. Higher Likert scores indicate images with superior annotation quality. For cases where the AI predictions were not rated as 'strongly agree', the reviewers corrected the AI annotations by editing the segmentations and saving the corrections to a NIFTI formatted file. These corrections were then used to calculate the quantitative accuracy of the AI models. To control project costs, the radiologist's review was limited to 10% of the entire dataset. The same 10% subset was also reviewed by the annotation specialist. For the remaining 90% of the data, the non-expert rated each AI prediction only on the Likert Scale. This allowed for an extrapolation of the correlation between the expert and non-expert ratings from the 10% subset to the remaining predictions. Thus, the radiologist and the annotation specialist both reviewed the same 10% of the data, ensuring overlap in their evaluations. The non-expert then rated the remaining 90% of the data independently.

The following sections describe the data curation, preprocessing, analysis, and results post-processing steps used to develop the AI models for each specified annotation task. The code to reproduce our analysis is publicly available via zenodo.org (Table 10).

*Table 2: Likert Score description used by reviewers to assess the quality of the AI annotations per case.*

| ***Likert Score*** | **Description** |
|---|---|
| **Strongly agree** | Use-as-is (i.e., clinically acceptable, and could be used for treatment without change) |
| **Agree** | Minor edits that are not necessary. Stylistic differences, but not clinically important. The current segmentation is acceptable. |
| **Neither agree nor disagree** | Minor edits that are necessary. Minor edits are those that the review judges can be made in less time than starting from scratch or are expected to have minimal effect on treatment outcome. |
| **Disagree** | Major edits. This category indicates that the necessary edit is required to ensure correctness, and sufficiently significant that user would prefer to start from the scratch. |
| **Strongly disagree** | Unusable. This category indicates that the quality of the automatic annotations is so bad that they are unusable. |

Volumetric segmentations produced by the models were saved as standard DICOM Segmentation objects (SEG), which included appropriate metadata to describe the contents and provide links to the input images. DICOM data element SegmentAlgorithmType (0062,0008) is set to "AUTOMATIC" if the segmentation is the AI output. If the segmentation is from a reviewer's correction, the SegmentAlgorithmType is set to "SEMIAUTOMATIC". The SegmentAlgorithmName (0062,0009) data element is set to a short name specific to the model, the specific value is given in the model overview sections below. The ContentCreatorName (0070,0084) data element and the SeriesDescription (0008,103E) data element contain the segmentation creator's description, such as AI, Radiologist, or Non-expert.

# FDG PET/CT Lung and Lung Tumor Annotation

## Imaging Data

***IDC Collections:*** TCGA-LUAD [3] , TCGA-LUSC [4] , LUNG-PET-CT-Dx [5], Anti-PD-1_Lung [6], RIDER Lung PET-CT [7], NSCLC-Radiogenomics [8–10], and ACRIN-NSCLC-FDG-PET [11,12] .

***Data Curation:*** For this AI-generated annotation task input images were attenuation-corrected paired FDG-PET/CT scans of the lung/chest region. Out of the seven chosen collections, a total of 736 paired FDG-PET/CT images matched the task criteria.

## Model Training Methodology

The AutoPET Challenge 2023 dataset [23,24] comprises whole-body FDG-PET/CT data from 900 patients, encompassing 1014 studies with tumor annotations. The highest performing model in the AutoPET II Challenge[25] used multitask learning by including tasks for organs that typically have background activity in FDG PET scans. This same multitask training strategy was employed by adding labels for the brain, bladder, kidneys, liver, stomach, spleen, lungs, and heart generated by the TotalSegmentator[22] model to the training dataset. A multi-task AI model was trained using the augmented datasets. To evaluate algorithm robustness and generalizability a held-out dataset of 150 studies, randomly selected without patient crossover, was employed. Among these, 100 studies were sourced from the same hospital as the training database, while 50 were selected from a different hospital but adhered to a similar acquisition protocol. This model has achieved robust results in the final leaderboard of AutoPET challenge [26,27]. The CT images were resampled to the resolution of the associated paired PET images.

## Annotation Data

***AI generated Annotations:*** The predictions of the AI-generated FDG-avid tumor annotation model[28] for this task were overlaid with the lung annotations provided by the TotalSegmentator[22] model. Tumor predictions were then limited to only the predictions seen in the pulmonary and pleural regions. An example output can be seen in Figure 3.

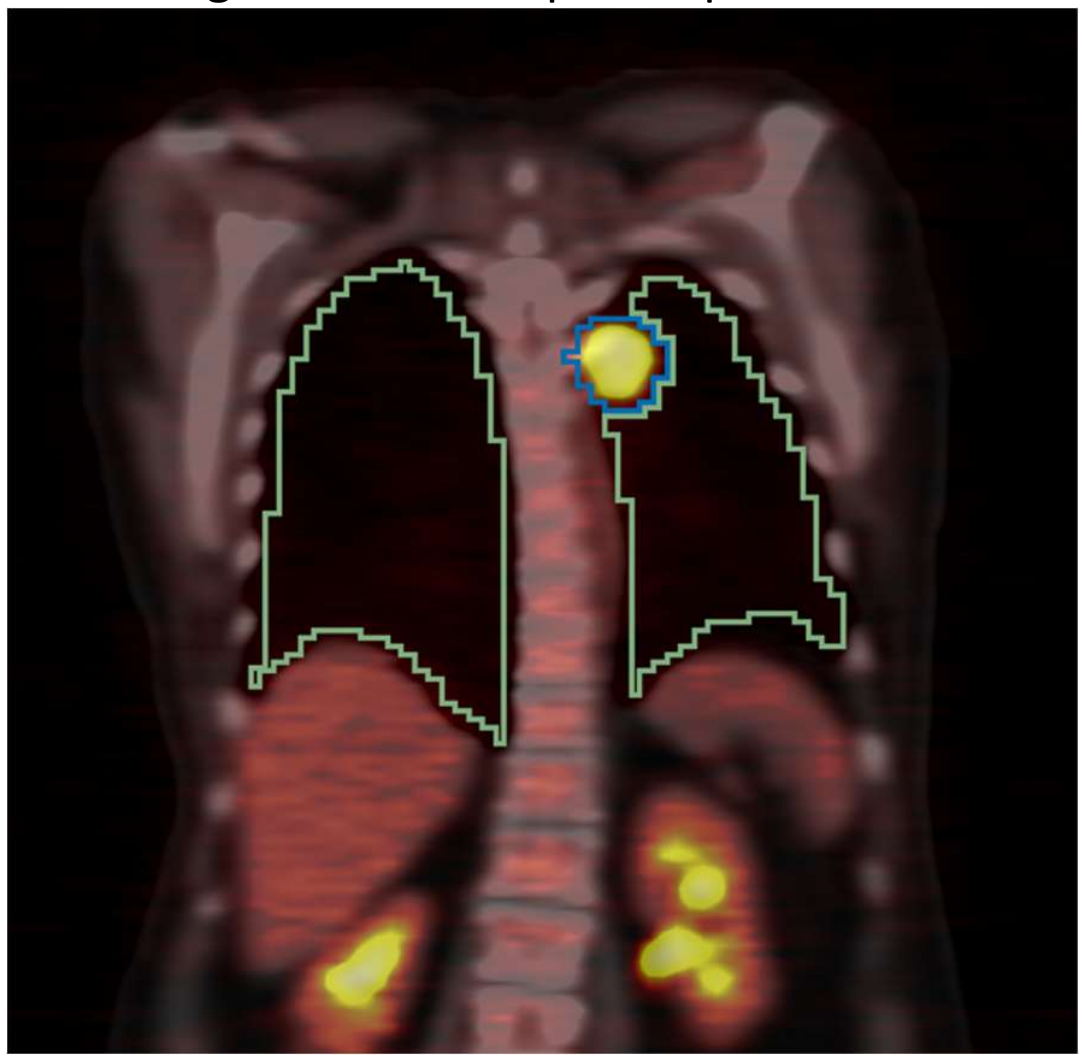

*Figure 3: Automatic segmentation of Lung (green) and FDG-avid tumor (blue) from FDG-PET/CT scans of patient RIDER-2610856938*

***Validation:*** The non-expert qualitatively assessed the AI annotations on all 736 DICOM studies using a Likert scale. Approximately 10% of the data (N=77) was randomly selected as a validation set. Both the non-expert and the expert Likert-scored and manually corrected the AI predictions of the validation set.

**DICOM-SEG SegmentAlgorithmType**: BAMF-Lung-FDG-PET-CT

# CT Lung Nodule Annotation

### Imaging Data

***IDC Collections:*** TCGA-LUAD [3] , TCGA-LUSC [4] , LUNG-PET-CT-Dx [5], Anti-PD-1_Lung [6], RIDER Lung PET-CT [7], and NSCLC-Radiogenomics[8–10]

***Data Curation:*** For this AI-generated annotation task input images were CT scans of the lung/chest region that were not part of the paired attenuation-corrected FDG-PET/CT scans that were used in the previous (FDG PET/CT Lung) task. Out of the six chosen collections, a total of 433 CT scans met the task criteria.

### Model Training Methodology

The DICOM-LIDC-IDRI-Nodules collection[29–31] was used to train an AI model[32,33] to annotate lung nodules. This collection included 883 studies with annotated nodules from 875 patients. Within the dataset only nodules that were identified by all four of their radiologists (size condition: 3mm ≤ diameter ≤ 30mm), were considered for AI model training for this task. The lung annotations AI model was trained on 411 and 111 lung CT data from NSCLC Radiomics[34,35] and NSCLC Radiogenomics[36] respectively. No additional preprocessing was used.

### Annotation Data

***AI generated Annotations :*** The predictions of the AI-generated lung nodule annotation model for this task were limited to by size and regions. Only annotations that were in the pulmonary and pleural regions and had diameters between 3mm and 30mm, same as training data, were kept. An example output can be seen in Figure 4.

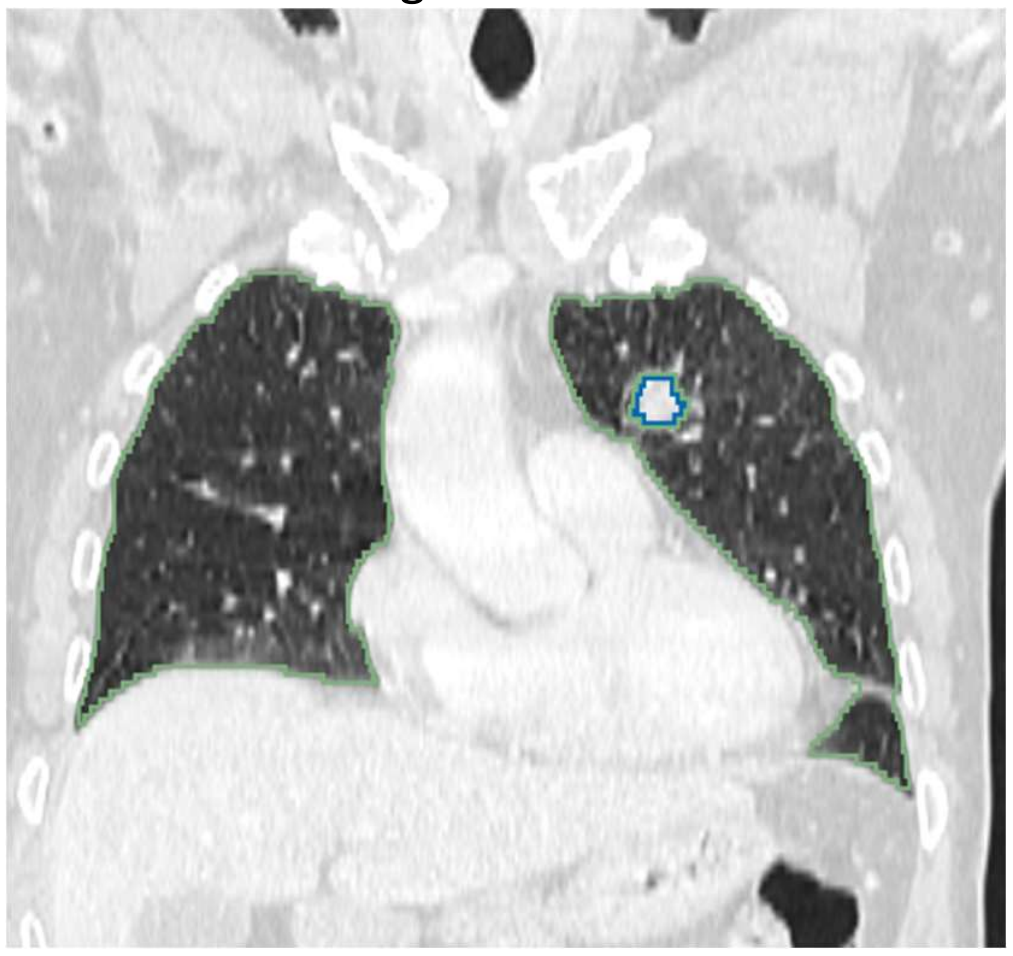

*Figure 4: Automatic segmentation of Lung (green) and nodule (blue) from CT scan of patient TCGA-34-5239*

***Validation:*** The non-expert qualitatively assessed all the AI annotations on all 430 DICOM studies using a Likert scale. Approximately 10% of the data (N=47) was randomly selected as a validation set. Both the non-expert and the expert Likert-scored and manually corrected the AI predictions of the validation set.

**DICOM-SEG SegmentAlgorithmType**: BAMF-Lung-CT

## FDG PET/CT Breast Tumor Annotation

### Imaging Data

***IDC Collections:*** QIN-Breast [13,14]

***Data Curation:*** For this AI-generated annotation input images were attenuation-corrected paired FDG-PET/CT. A total of 110 paired PET/CT scans met the task criteria.

### Model Training Methodology

***Model design and training:*** This task used the same nnU-Net[20] based AI model[28] as the previous FDG-PET/CT Lung and FDG-avid Tumor tasks, which was trained on the AutoPET Challenge 2023 dataset augmented for multitask by incorporating labels generated by TotalSegmentator[22]. The CT images were resampled to the resolution of the paired PET images.

### Annotation Data

***AI generated Annotations :*** The predictions of the AI-generated FDG-avid tumor annotation model for this task were overlayed with the annotations provided by the TotalSegmentator[22] model. Tumor predictions were then limited to only the predictions seen in the breast regions. An example output can be seen in Figure 5.

***Validation:*** The non-expert qualitatively assessed the AI annotations on all 110 DICOM studies using a Likert scale. Approximately 10% of the data (N=10) was randomly selected as a validation set. Both the non-expert and the expert Likert-scored and manually corrected the AI predictions of the validation set.

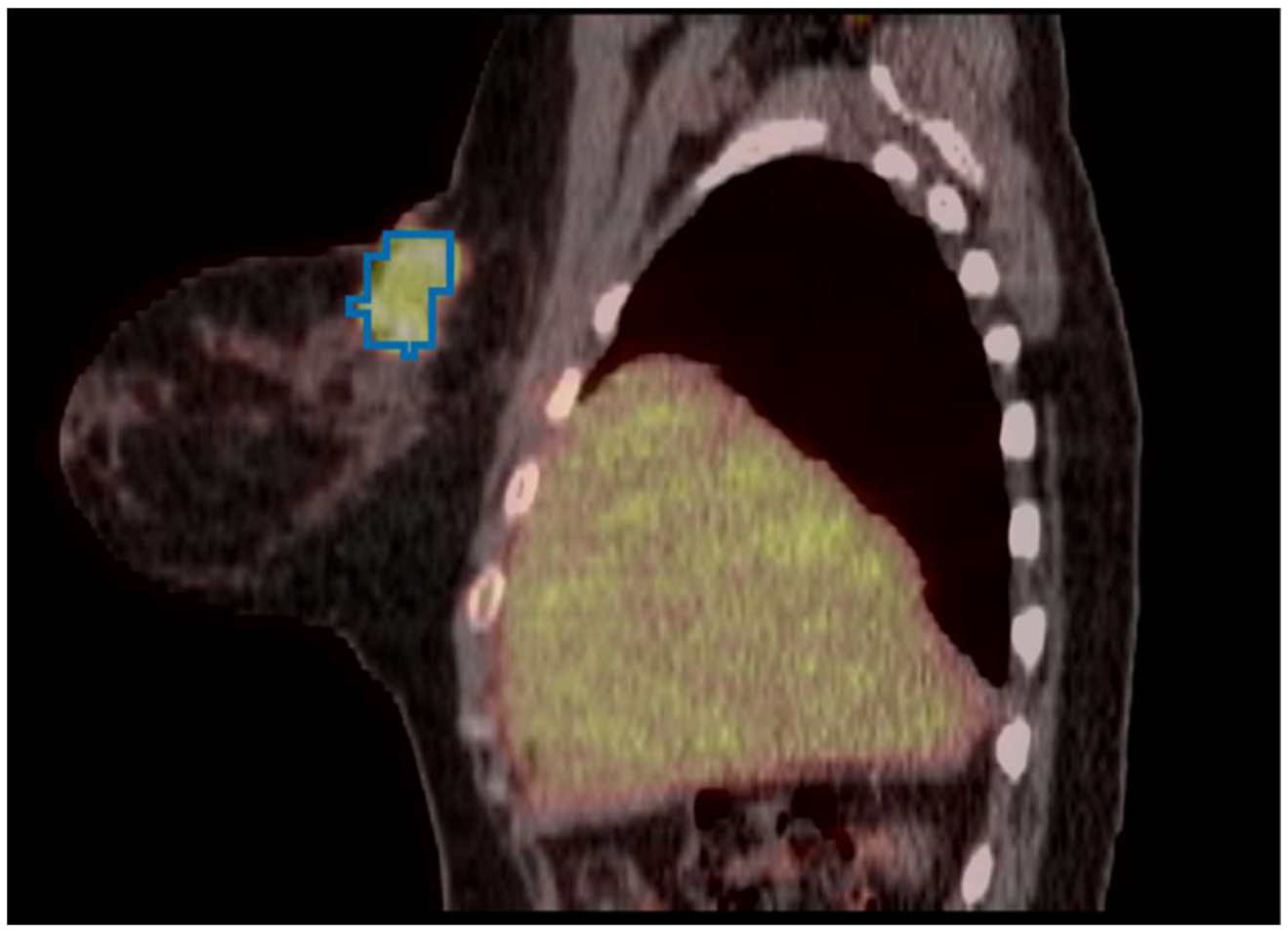

*Figure 5: Automatic segmentation of FDG-avid breast tumor (blue) from FDG-PET/CT scans of patient QIN-BREAST-01-0033*

**DICOM-SEG SegmentAlgorithmType**: BAMF-Breast-FDG-PET-CT

## CT Kidneys, Tumors, and Cysts Annotation

Imaging Data

***IDC Collections:*** TCGA-KIRC [15]

***Data Curation:*** For this AI-generated annotation task, input images were limited to contrast enhanced CT scans that contained the kidneys. A total of 156 CT scans met the task criteria.

### Model Training Methodology

The kidney tumor annotation AI model was trained to accurately delineate the kidney, tumor, and cysts. Model training was split into two stages. Stage one training used contrast CTs from the KiTS 2021 collection[37–39] (N=489) to identify the kidney, tumor, and cysts. This trained model was then used to generate annotations for 64 cases of TCGA-KIRC[15] collection. These annotations were then further refined by non-experts. An additional 45 cases from the TCGA-KIRC [15] dataset was included as part of the training set for stage two training. The final trained model[40] was used to generate annotations for all 156 cases of the TCGA-KIRC [15] collection that met the task criteria. No additional preprocessing was used.

### Annotation Data

***AI generated Annotations:*** The AI-generated annotations were limited to the two largest connected components to remove false positives. The connected components were determined from the union of the kidney, cyst, and tumor labels. An example output can be seen in Figure 6.

***Validation:*** The non-expert qualitatively assessed the AI annotations on all 156 DICOM studies using a Likert scale. Approximately 20% of the data (N = 39) was randomly selected as a validation set. A larger percentage of the collection was selected for annotation because of the heterogeneity from characteristics such as contrast phase and scan field of view. Both the non-expert and the expert Likert-scored and manually corrected the AI predictions of the validation set. Additionally, the expert provided annotations for 39 cases. This enabled a comparison between the final model's annotations and expert's annotations.

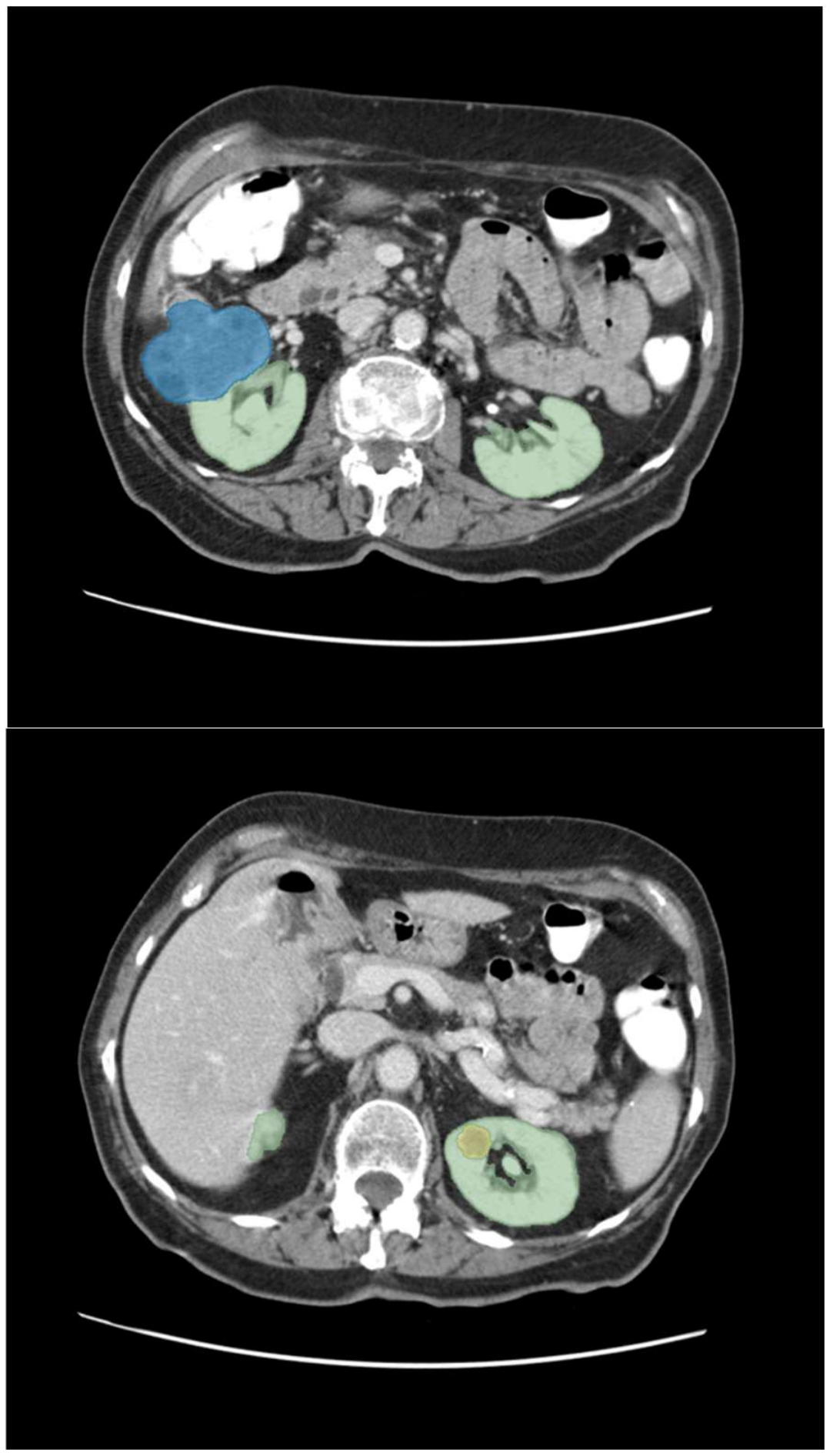

*Figure 6: Automatic segmentation of kidney (green), tumor (blue), cyst (yellow) from CT scan of patient TCGA-CJ-4873*

**DICOM-SEG SegmentAlgorithmType**: BAMF-Kidney-CT

## MRI Prostate Annotation

### Imaging Data

***IDC Collections:*** ProstateX [16,17]. For this collection, the IDC has prostate annotations for 98 MRI scans from PROSTATEx-Seg-HiRes[41,42] (high resolution prostate annotations, N = 66) and PROSTATEx-Seg-Zones[43,44] (zone segmentations of the prostate, N = 32).

***Data Curation:*** For this AI-generated annotation task, input images were limited to T2W MRI scans that did not have any missing slices. A total of 347 MRI scans met the task criteria.

### Model Training Methodology

***Model design and training:*** While an extensively trained prostate AI segmentation model already exists in the PI-CAI collections[45] dataset (N = 1500) we were unable to use this model as a baseline due to their inclusion of the ProstateX [16,17] dataset for their training and validation. To ensure no ProstateX [16,17] data leakage occurred in the AI model training this task was done in two stages. The training set for the first stage modal was composed of manual annotations from several datasets. It included 232 scans from ProstateX [16,17], where 98 of the labels were in the IDC collection and 134 were from the PROSTATEx_masks[46,47] collection. An additional 207 data points came from Prostate158[48] (N = 138) and ISBI-MR-Prostate-2013[49] (N = 69). Both of these datasets contained a single case that did not meet our inclusion criteria and thus that case was excluded. A total of 439 T2W MRI prostate annotations were used to train the first stage prostate annotation AI model. A test/holdout validation split of 81/34 was created from the remaining 115 scans without annotations in the ProstateX [16,17] collection. The first stage model was then used to predict the unannotated 81 test set scans of ProstateX [16,17] and 1172 cases of the PI-CAI collections[45]. All ProstateX [16,17] scans including the same patient were removed from the PI-CAI dataset (N = 1500) to ensure no data leakage between the two collections. A portion of the PI-CAI dataset contained a much larger field of view than the field of view used in the training collections. To combat the increased risk of additional off-targeting regions in the predictions the centremost segmentation (in all directions) was assumed to be the prostate and all additional regions were removed for all 1253 prostate predictions. In the second training stage a new AI model[50] was trained using the same data as the first stage but now with the addition of the 1253 predicted annotations from the ProstateX [16,17] test split (N=81) and the PI-CAI collections (N=1172). This second stage was used to generate the final prostate labels for the ProstateX [16,17] collection. The 34 scans from the ProstateX [16,17] holdout collection was used for manual validation by radiologist and non-expert. No additional preprocessing was used.

### Annotation Data

***AI generated Annotations :*** The AI-generated prostate annotations were limited to the largest centremost (in all directions) annotation. An example output can be seen in Figure 7.

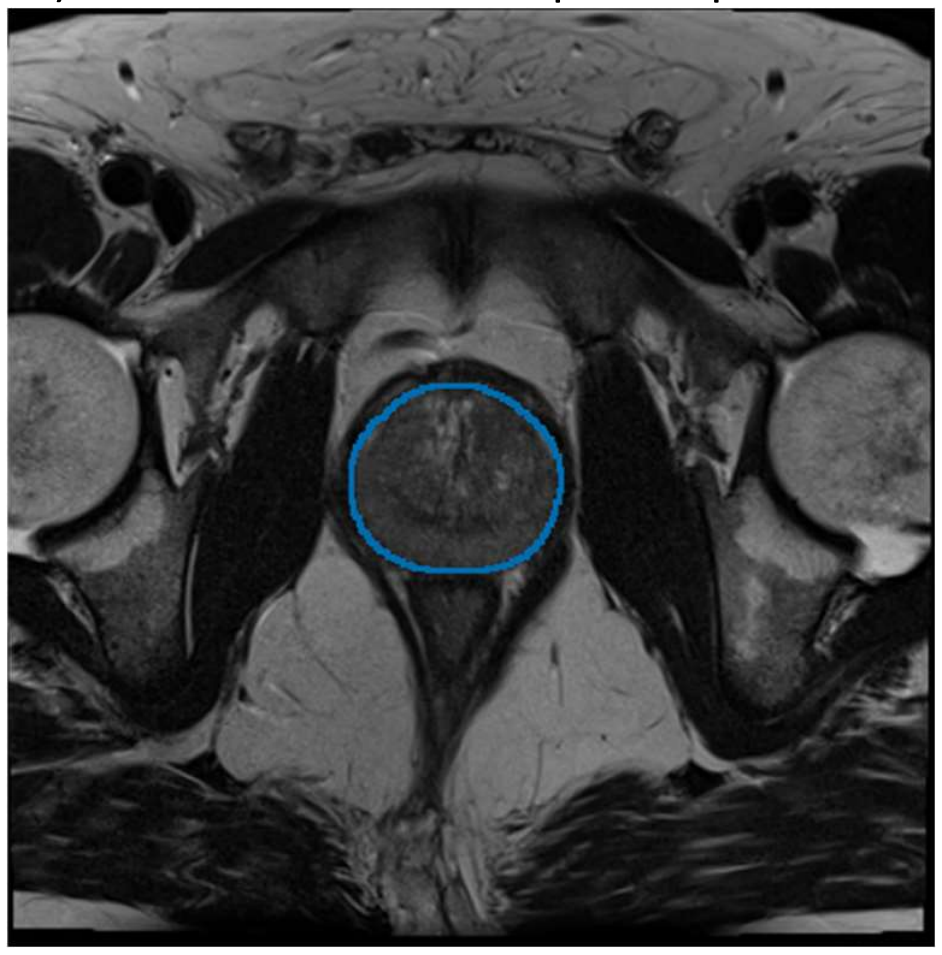

*Figure 7: Automatic segmentation of prostate gland from T2 MRI scan of patient ProstateX-0336*

***Validation:*** The non-expert qualitatively assessed the AI annotations on all 347 DICOM studies using a Likert scale. Approximately 10% of the data (N=34) was randomly selected as a validation set. Both the non-expert and the expert Likert-scored and manually corrected the AI predictions of the validation set. Additional validation was performed by generating AI segmentation for the QIN-Prostate-Repeatability[51–55], PROMISE12[56], and Medical Segmentation Decathlon[57] T2W MRI collections.

**DICOM-SEG SegmentAlgorithmType**: BAMF-Prostate-MR

## MRI Liver Annotation

### Imaging Data

***IDC Collections:*** TCGA-LIHC [18]

***Data Curation:*** For this AI-generated annotation task, input images were limited to T21W MRI. A total of 65 MRI scans met the task criteria.

### Model Training Methodology

350 MRI liver annotations taken from the AMOS[58] (N=40) and DUKE Liver Dataset V2[59] (N=310) collections were used to train an MRI liver annotation AI model[60]. No additional preprocessing was used.

### Annotation Data

***AI generated Annotations :*** The AI-generated liver annotations were limited to the single largest connected component. An example output can be seen in Figure 8.

***Validation:*** A non-expert qualitatively assessed the AI annotations on all 65 DICOM studies using a Likert scale. Approximately 10% of the data (N=7) was randomly selected as a validation set. Both the non-expert and the expert Likert-scored and manually corrected the AI predictions of the validation set.

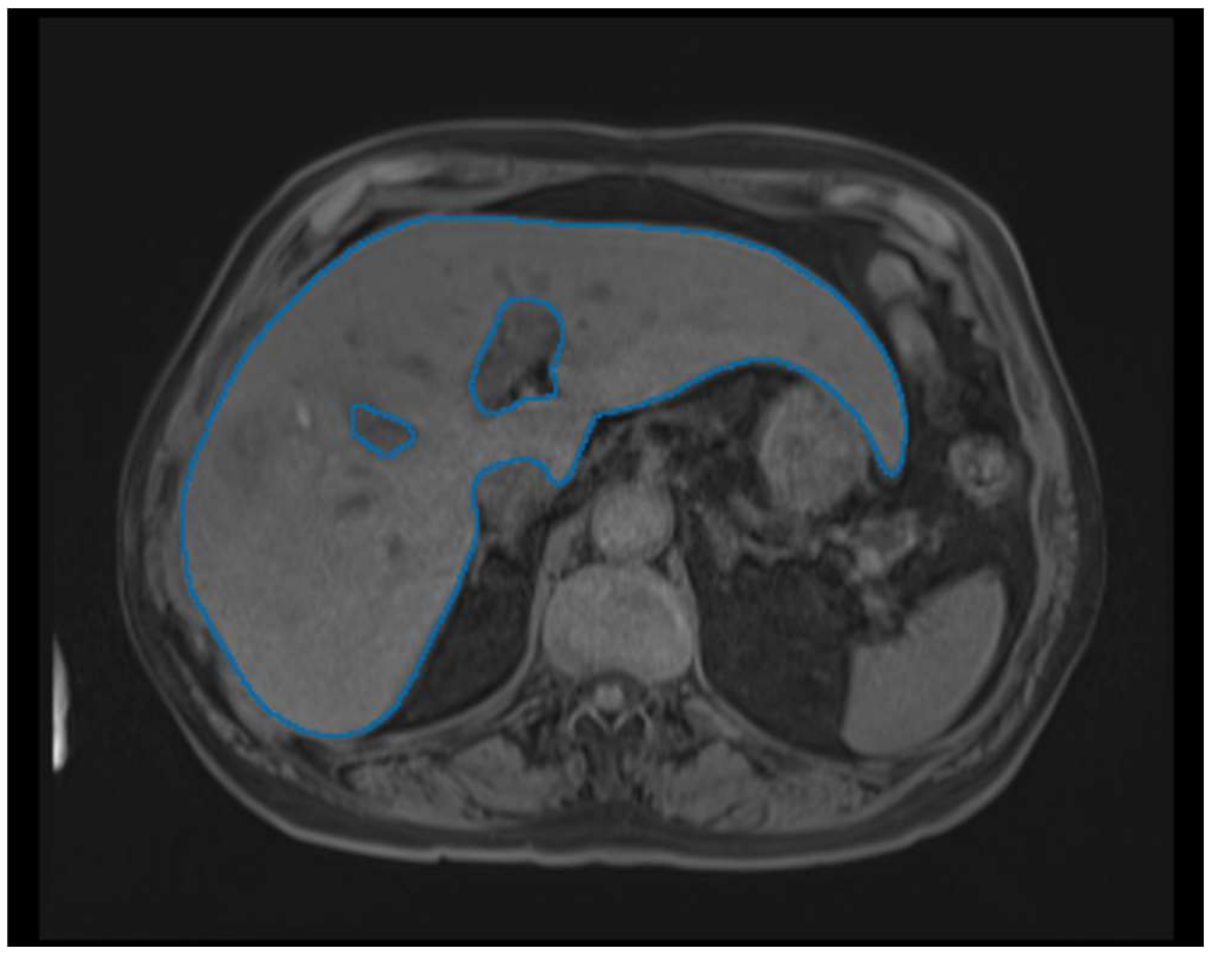

*Figure 8: Automatic segmentation of the liver from T1 MRI scan of patient TCGA-G3-A7M7*

**DICOM-SEG SegmentAlgorithmType**: BAMF-Liver-MR

## CT Liver Annotation

### Imaging Data

***IDC Collections:*** TCGA-LIHC [18]

***Data Curation:*** For this AI-generated annotation task, input images were limited to CT scans of the liver region. A total of 89 CT scans met the task criteria.

### Model Training Methodology

1565 CT liver annotations taken from the TotalSegmentator[22] (N=1204) and FLARE21[61,62] (N=361) collections were used to train a CT liver annotation AI model[63]. No additional preprocessing was used.

### Annotation Data

***AI generated Annotations :*** The AI-generated liver annotations were limited to the single largest connected component. An example output can be seen in Figure 9.

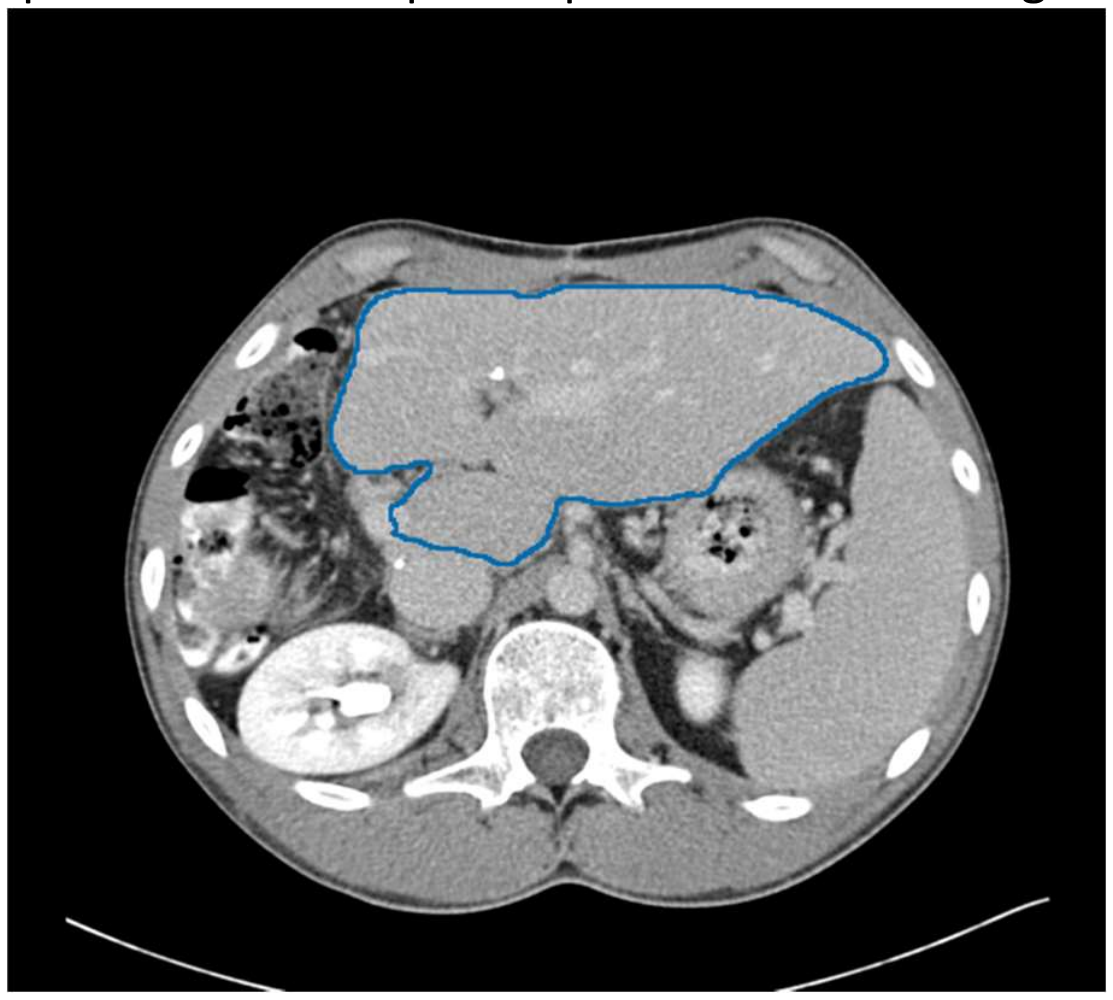

*Figure 9: Automatic segmentation of the liver from CT scan of patient TCGA-DD-A1EH*

***Validation:*** A non-expert qualitatively assessed all liver annotations using a Likert scale. Approximately 10% of the data (N=9) was randomly selected as a validation set. Both the non-expert and the expert Likert-scored and manually corrected the AI predictions of the validation set.
**DICOM-SEG SegmentAlgorithmType**: BAMF-Liver-CT

## Data Records

The reviewers scoring and comments, as well as DICOM Segmentation objects for the AI predictions and reviewer's corrections are available in Zenodo[64] (https://doi.org/10.5281/zenodo.13244892).

Each zip file in the collection correlates to a specific segmentation task. The common folder structure is:

- *ai-segmentations-dcm* This directory contains the AI model predictions in DICOM-SEG format for all analysed IDC collection files.
- *qa-segmentations-dcm* This directory contains manual corrected segmentation files, based on the AI prediction, in DICOM-SEG format. Only a fraction, ~10%, of the AI predictions were corrected. Corrections were performed by radiologist (rad*) and non-experts (ne*).
- *qa-results.csv* CSV file linking the study/series UIDs with the ai segmentation file, radiologist corrected segmentation file, radiologist ratings of AI performance. Reviewer Likert scores and review comments for the segmentations are also included in this file.

The DICOM segmentation files can also be linked back to the original DICOM scans from the IDC using DICOM metadata in each segmentation file. The DICOM data element SeriesInstanceUID (0020,000E) in the data element ReferencedSeriesSequence (0008,1115) refers back to the original DICOM scan this segmentation was derived from.

The DICOM segmentations have been integrated into the IDC. From that portal it is possible to view the segmentations overlayed on the images they were derived from. The direct link to the segmentation collection is in IDC (https://portal.imaging.datacommons.cancer.gov/explore/filters/?analysis_results_id=BAMF-AIMI-Annotations)

## Technical Validation

The AI models were evaluated on the following series of metrics. Some of these were only applicable to a subset of the model tasks.

- Kendall's $\tau$: measure the correlation between ordinal data, in this case, the Likert scores of the manual reviewers.
- Sørensen–Dice coefficient[65] (DSC): measures the similarity between volumetric segmentations, $V_A$ and $V_B$. It is twice the intersection of the volumes over the sum of the volumes.

$$DSC = \frac{2(V_A \cap V_B)}{V_A + V_B}$$

- Normalized Surface Dice[66] (NSD): measures surface distance similarity. It measures the amount of the surface of a volume ($S_A$) that is within a tolerance ($\tau$) of the surface of another volume ($S_B^\tau$). This is calculated for both surfaces and normalized to the total surface of the volumes. The tolerance distance is task specific and are from the Medical Segmentation Decathlon[57].

$$NSD = \frac{(S_A \cap S_B^\tau) + (S_B \cap S_A^\tau)}{S_A + S_B}$$

- 95% Hausdorff Distance: measures surface agreement. It is the distance at which 95% of the points on Surface A have a point on Surface B less than it.

For models with tumor prediction, we can measure the detection rate as follows:

- True positive: the predicted tumor overlaps with a ground truth tumor
- False positive: the predicted tumor does not overlap with a ground truth tumor
- False negative: the ground truth tumor does not overlap with a predicted tumor

From these, tumor detection sensitivity, false negative rate, and F1 score were calculated.

The following sections in the results are organized in the same order that was used in the methods section.

### FDG PET/CT Lung and Lung Tumor Annotation

The 77 validation cases were rated by a radiologist and non-expert. Most cases were rated 'Strongly Agree', meaning no changes were required (Figure 10). This accounts for the good mean metrics shown in Table 3 although the lower scoring tickets cause a high standard deviation in the metrics. An important note is that there is no significant correlation as shown by Kendall's $\tau$. While both reviewers agreed that most of the cases should be scored 'Strongly Agree', they disagreed on which cases those were. (https://doi.org/10.5281/zenodo.13851020).

*Table 3: Label-wise segmentation metrics (mean (standard deviation)) between AI derived and manual corrected FDG PET/CT lungs and tumor annotations. Detection Accuracy is given between the AI and the 'ground truth' of the expert's corrections.*

| | **Expert** | | **Non-Expert** | |
|---|---|---|---|---|
| ***Segmentation Metric*** | **Lung** | **Tumor** | **Lung** | **Tumor** |
| **DSC** | 1.00 (±0.00) | 0.97 (±0.11) | 0.99 (±0.04) | 0.92 (±0.20) |
| **95% HD (mm)** | 0.10 (±0.58) | 5.83 (±19.42) | 1.97 (±10.50) | 10.00 (±26.34) |
| | | | | |
| ***Detection Accuracy*** | | | | |
| **Sensitivity** | | 0.91 | | |
| **False negative rate** | | 0.09 | | |
| **F1 score** | | 0.94 | | |

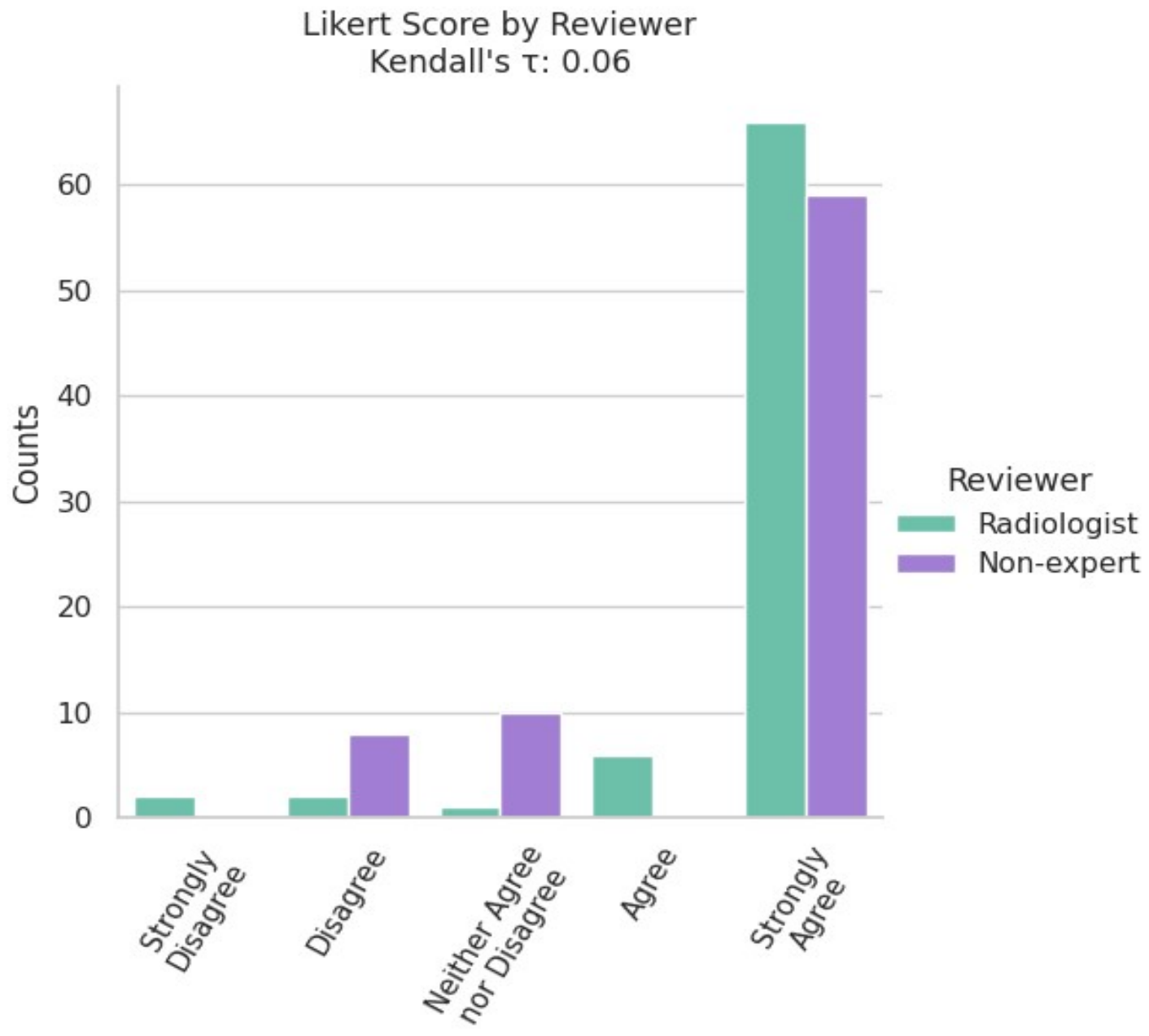

*Figure 10: Counts of Likert Scores for validation set of Lung PET/CT Lung Tumor model by reviewers*

### CT Lung Nodule Annotation

A validation set of 47 cases were reviewed by two radiologists. These reviewers had a moderate correlation of the Likert Scores (Kendall's $\tau$ = 0.42) The variety of Likert Scores show a mix of quality output for the model, Figure 11. The model performance metrics for the validated cases are listed in Table 4. Segmentation metrics of the lung label is very high, but lower and variable for the nodule label. These lower segmentation metrics are also reflected by the low nodule detection accuracy of the model. (https://doi.org/10.5281/zenodo.13851336).

*Table 4: Label-wise metrics (mean (standard deviation)) between AI derived and expert corrected CT lungs and nodules annotations. Detection Accuracy is given between the AI and the 'ground truth' of the expert's corrections.*

| | **Expert 1** | | **Expert 2** | |
|---|---|---|---|---|
| ***Segmentation Metric*** | **Lung** | **Nodule** | **Lung** | **Nodule** |
| **DSC** | 0.99 (±0.02) | 0.60 (±0.42) | 1.00 (±0.00) | 0.78 (±0.34) |
| **95% HD (mm)** | 2.34 (±5.89) | 56.72 (±64.36) | 0.30 (±1.70) | 26.06 (±48.63) |

| ***Detection Accuracy*** | **Expert1** | **Expert 2** |
|---|---|---|
| **Sensitivity** | 0.26 | 0.37 |
| **False negative rate** | 0.74 | 0.63 |
| **F1 score** | 0.41 | 0.54 |

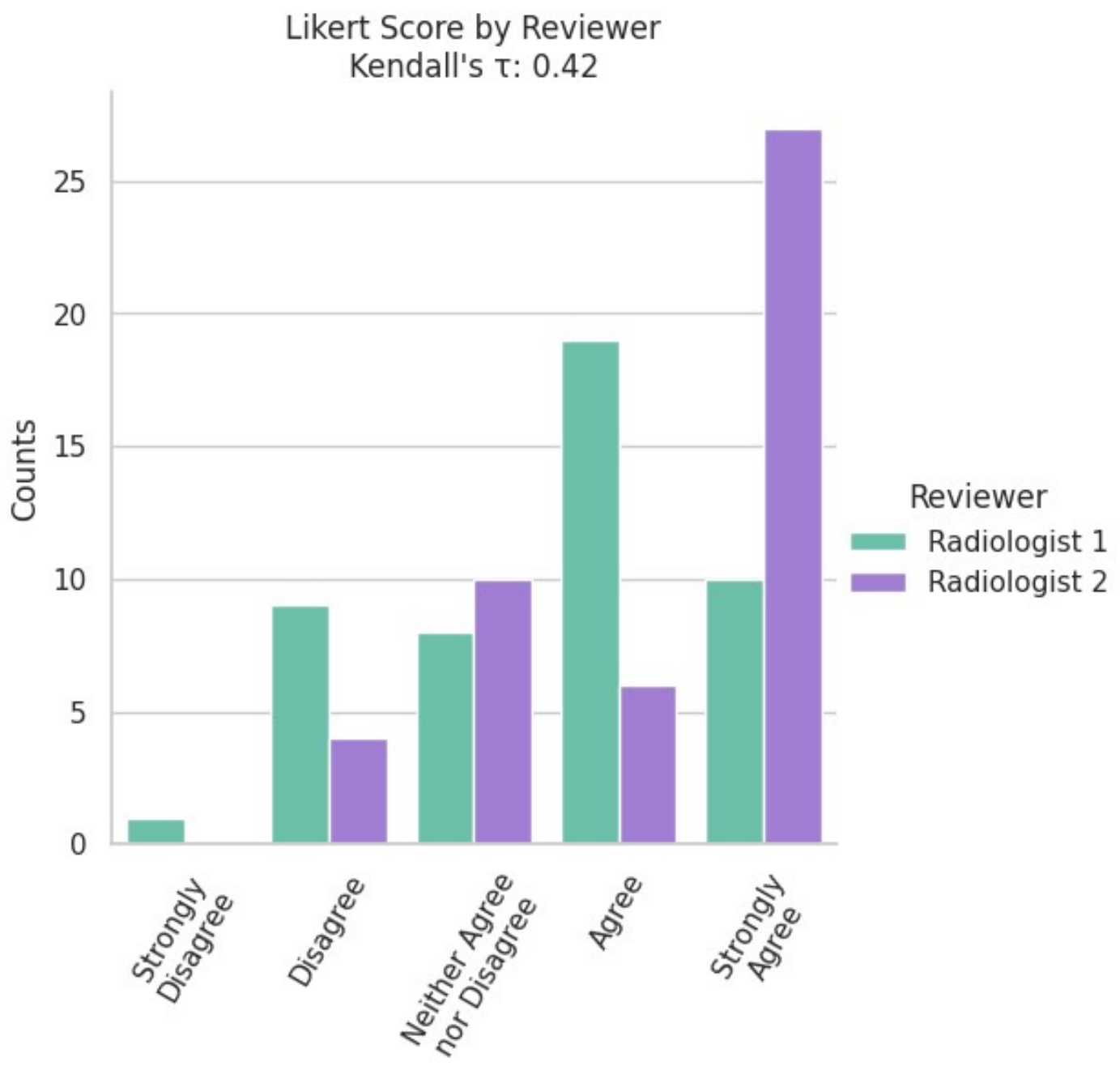

*Figure 11: Counts of Likert Scores for validation set of Lung CT Lung Nodule model by reviewers.*

### FDG PET/CT Breast Tumor Annotation

A radiologist and non-expert reviewed the 11 validation set cases for the FDG PET/CT Breast tumor model. There was moderate correlation between reviewers Likert scores, Kendall's $\tau$=0.58, shown in Figure 12. Generally, the model did well as shown by the performance metrics in Table 5. The radiologist commented that the AI's most common failure was to include FDG-avid retropectoral lymph nodes in the breast tumor label. (https://doi.org/10.5281/zenodo.13851345).

*Table 5: Label-wise metrics (mean (standard deviation)) between AI derived and expert corrected FDG PET/CT breast lesion annotations. Detection Accuracy is given between the AI and the 'ground truth' of the expert's corrections.*

| | Expert 1 | Non-Expert |
|---|---|---|
| ***Segmentation Metric*** | **Tumor** | **Tumor** |
| **DSC** | 0.80 (±0.33) | 0.94 (±0.10) |
| **95% HD (mm)** | 29.70 (±33.43) | 13.53 (±20.00) |
| | | |
| ***Detection Accuracy*** | | |
| **Sensitivity** | **0.43** | |
| **False negative rate** | **0.57** | |
| **F1 score** | **0.52** | |

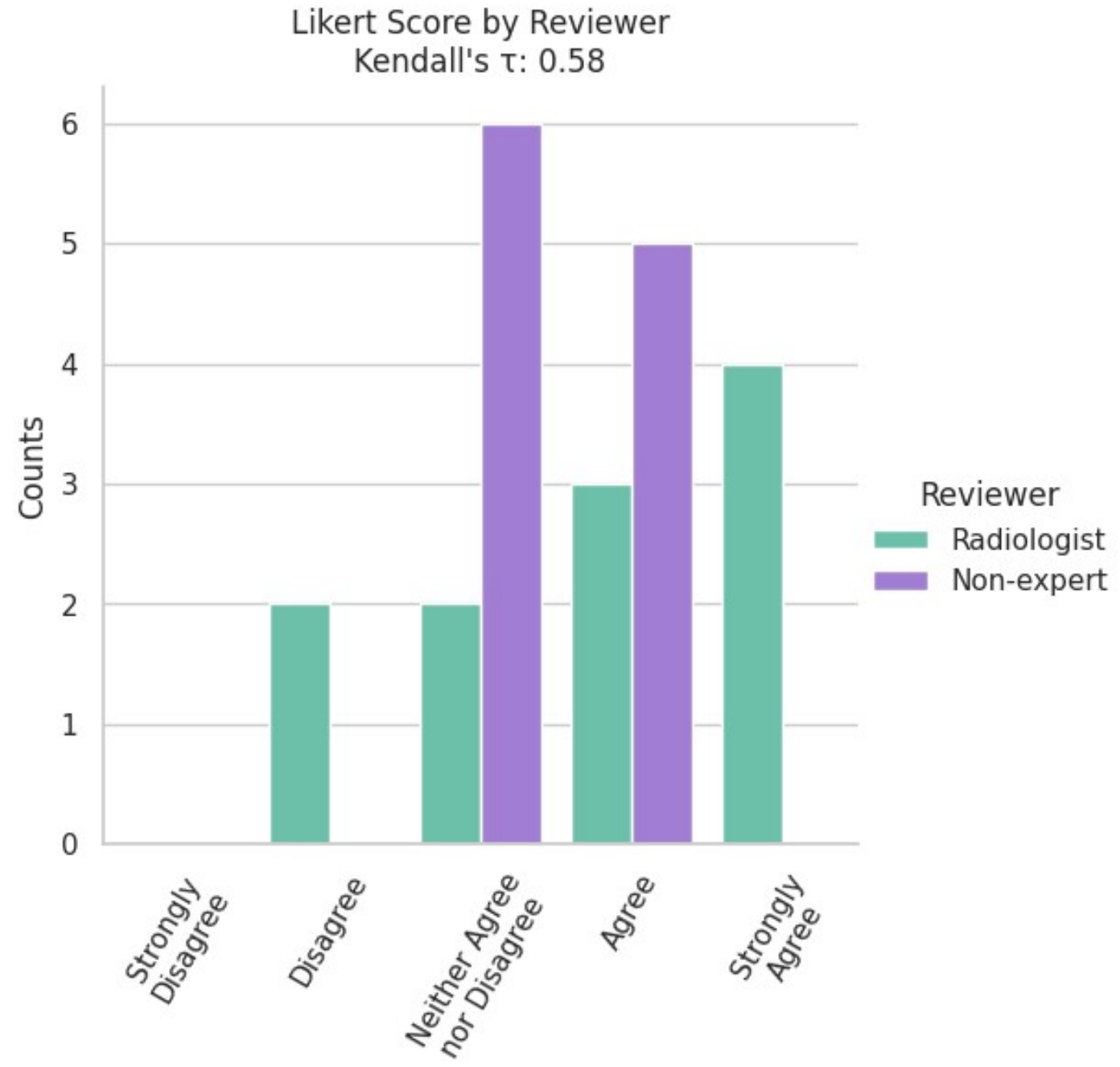

*Figure 12: Counts of Likert Scores for validation set of Breast FDG-avid tumor model by reviewers.*

## CT Kidneys, Tumors, and Cysts Annotation

The validation set comprised 39 cases from the TCGA-KIRC [15] IDC collection. Since the model was trained on KiTS23 [37–39] data, the validation set was evaluated using the same metrics as the KiTS23[37–39] challenge. The Dice coefficient and Normalized Surface Distance were calculated for three combinations of labels: Kidney+Tumors+Cysts, Tumors+Cysts, Tumors. These metrics are shown in Table 6. Additionally, two manual reviews, a radiologist and nonexpert rated the validation set on a Likert Scale and had moderate correlation in scores as seen in Figure 13. The TCGA-KIRC [15] collection contains scans with a variety of contrast phases. The reviewers manually identified the contrast phase of the scan. Figure 14 shows that mean Likert scores were higher for scans in the nephrogenic and corticomedullary phase than those in the excretory phase or without contrast. The higher performance of the nephrogenic and corticomedullary phase was expected because the model training set contained data from only these phases. It was interesting that scans from the Excretory phase performed as well considering the absence of this phase from the training data. (https://doi.org/10.5281/zenodo.13851351).

*Table 6: Model performance on TCGA-KIRC data as reviewed by Radiologist and Non-expert. Metrics are in mean (standard deviation) format.*

| Metric | Expert 1 | | | Non-Expert | | |
|---|---|---|---|---|---|---|
| *Metric* | Kidneys Tumor Cyst | Tumor Cyst | Tumor | Kidneys Tumor Cyst | Tumor Cysts | Tumor |
| DSC | 0.93 (±0.22) | 0.88 (±0.32) | 0.88 (±0.32) | 0.99 (±0.06) | 0.96 (±0.19) | 0.96 (±0.19) |
| NSD | 0.91 (±0.23) | 0.87 (±0.32) | 0.87 (±0.32) | 0.98 (±0.09) | 0.95 (±0.2) | 0.95 (±0.2) |
| 95% HD (mm) | 13.60 (±45.58) | 7.06 (±23.88) | 7.10 (±24.21) | 3.43 (±17.64) | 0.83 (±5.02) | 0.83 (±5.05) |

| *Detection Accuracy* | Tumor | Cysts |
|---|---|---|
| **Sensitivity** | 0.88 | 0.94 |
| **False Negative Rate** | 0.12 | 0.06 |
| **F1 score** | 0.91 | 0.96 |

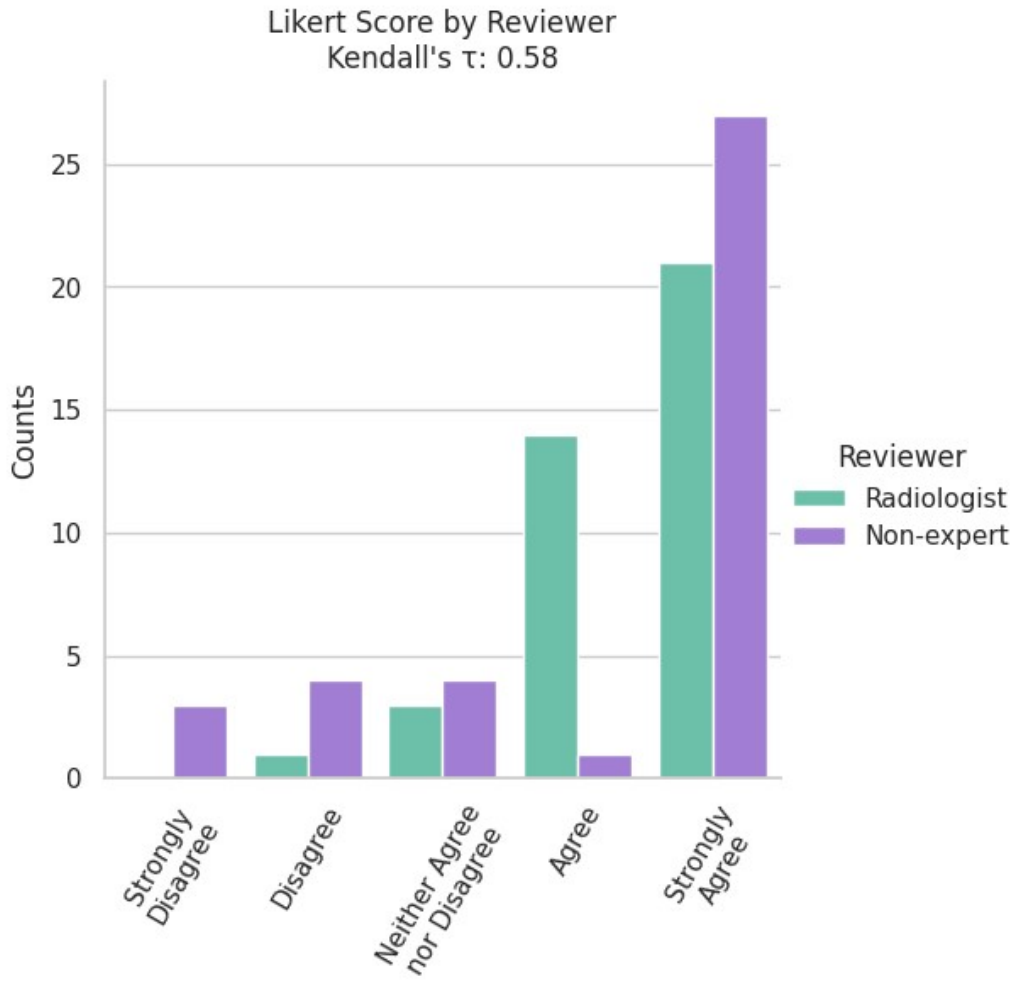

*Figure 13: Counts of Likert Scores for validation set of Kidneys CT model by reviewers.*

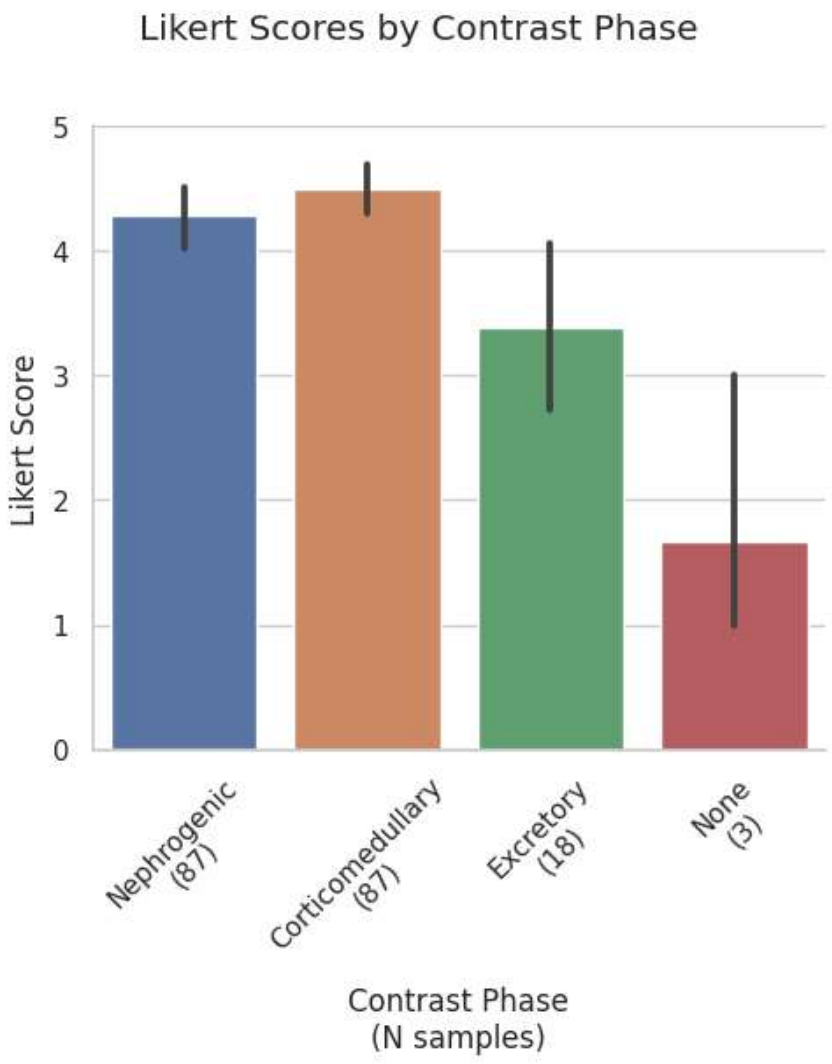

*Figure 14: Mean Likert scores by contrast phase for all AI predictions from both Radiologist and non-expert.*

## MRI Prostate Annotation

A radiologist and non-expert reviewed the 34 validation set cases for the MRI prostate segmentation model. Both reviewers rated all 34 cases 'Strongly Agree', Figure 16. In addition to the 34 case validation set from ProstateX [16,17], the model's performance was also measured against other collections with existing prostate gland segmentation, QIN-Prostate, PROMISE12, and the Medical Segmentation Decathlon (MSD). These metrics are in Table 7, The Dice Coefficient distribution is shown in Figure 15. For this model, we used the NSD with a tolerance of 4mm as was done by the Medical Segmentation Decathlon. (https://doi.org/10.5281/zenodo.13851368).

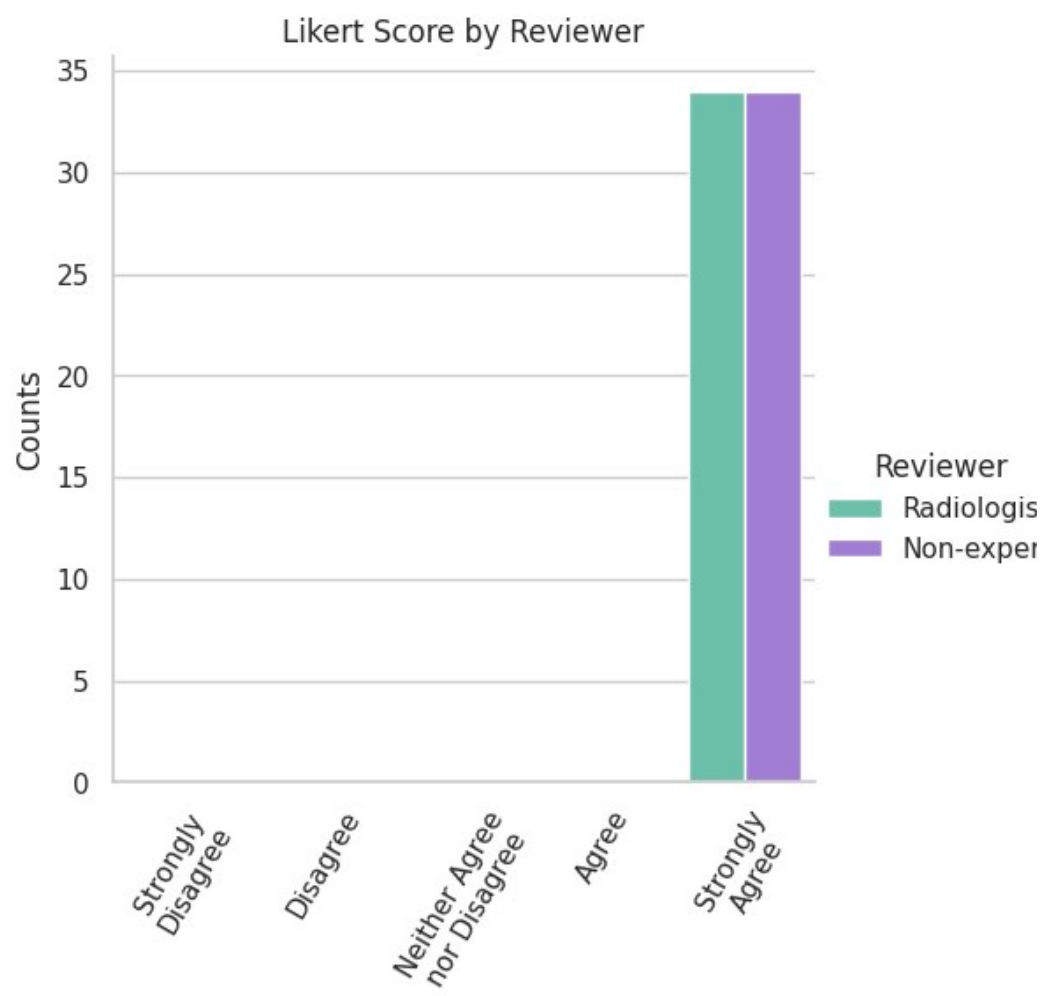

*Figure 15: Counts of Likert Scores for ProstateX validation set of MRI Prostate model by reviewers.*

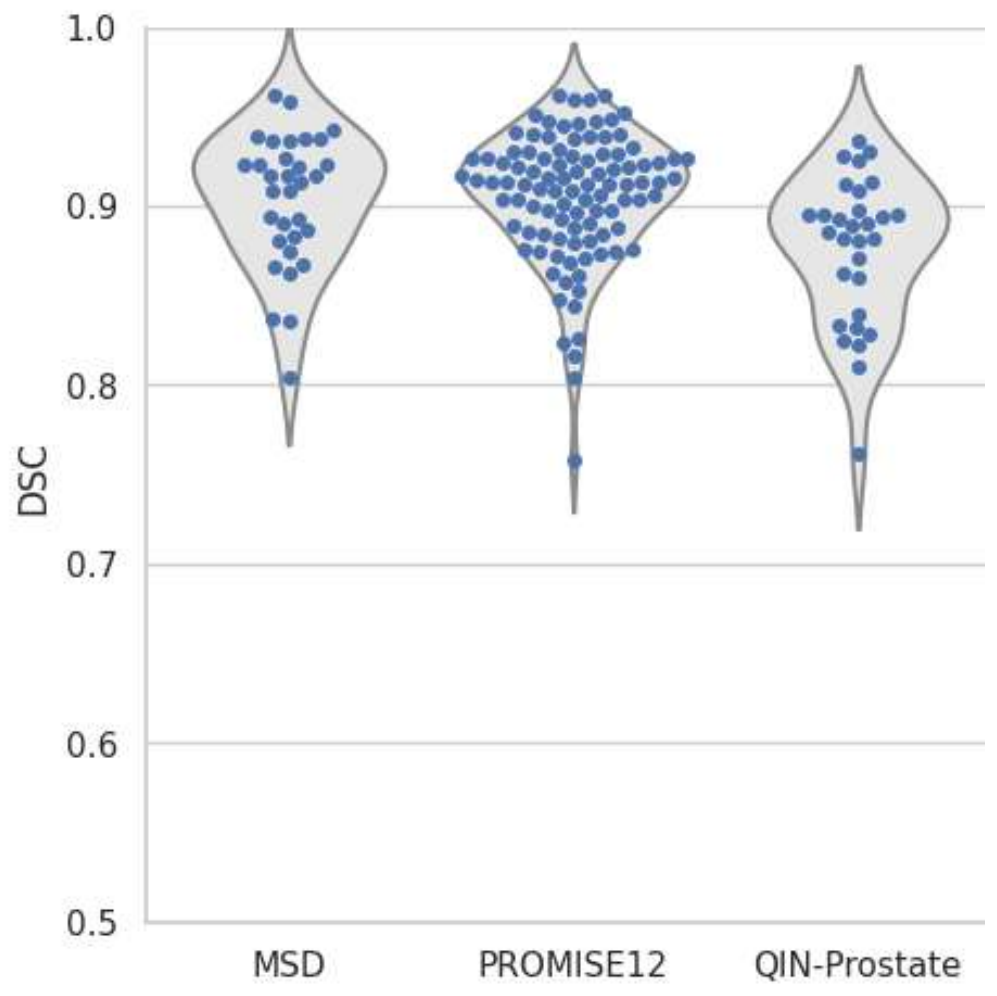

*Figure 16: Dice Coefficient of AI prostate gland segmentation for Medical Segmentation Decathlon, PROMISE12, and QIN-Prostate datasets.*

*Table 7: Label-wise metrics (mean (standard deviation)) between AI derived and the 'ground truth' expert corrected MRI Prostate annotations and other publicly available prostate annotation datasets.*

| ***Segmentation Metric*** | **IDC ProstateX** | **MSD** | **PROMISE12** | **QIN-Prostate** |
|---|---|---|---|---|
| **DSC** | 1.00 (±0.00) | 0.90 (±0.04) | 0.91 (±0.04) | 0.88 (±0.04) |
| **$NSD_4$** | 0.00 (±0.00) | 0.97 (±0.04) | 0.97 (±0.04) | 0.95 (±0.06) |
| **95% HD (mm)** | 0.00 (±0.00) | 3.65(±0.98) | 4.10(±1.93) | 6.06(±4.62) |

## MRI Liver Annotation

A set of 7 cases from the TCGA-LIHC [18] collection was selected as a validation set for the MRI liver segmentation model. A radiologist and non-expert reviewed these 7 cases. The reviewers had a high correlation in the Likert scores as shown in Figure 17. Table 8 shows the DSC and NSD metrics for the validation set. A tolerance of 7mm was used for the NSD metric, same as the CT liver segmentation task from the Medical Segmentation Decathlon. (https://doi.org/10.5281/zenodo.13851371).

*Table 8: Label-wise metrics (mean (standard deviation)) between AI derived and the 'ground truth' expert corrected MRI liver annotations.*

| **Metric** | **Expert1** | **Non-expert** |
|---|---|---|
| **DSC** | 0.91 (±0.18) | 0.90 (±0.15) |
| **NSD** | 0.89 (±0.20) | 0.85 (±0.20) |
| **95% HD (mm)** | 14.66(±25.98) | 33.21(±47.96) |

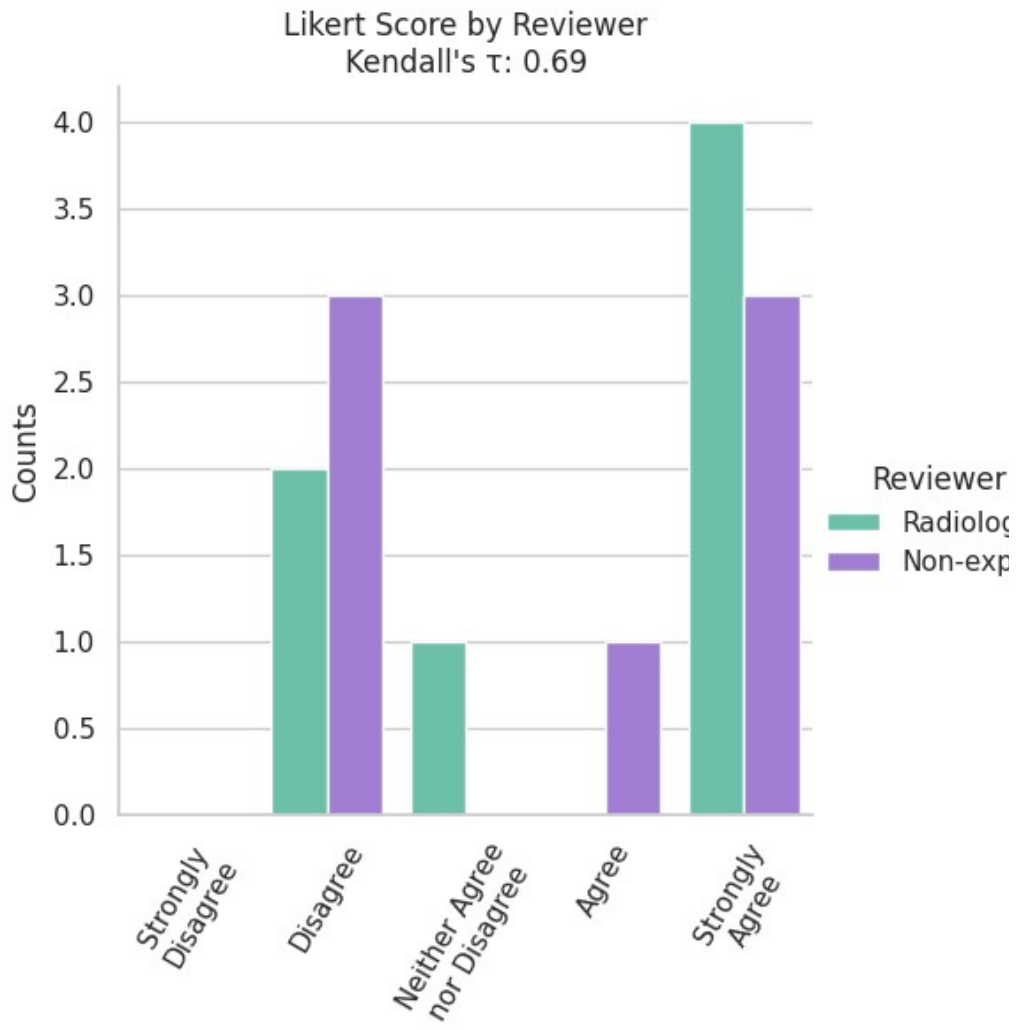

*Figure 17: Reviewer Likert score comparison for MRI Liver segmentations.*

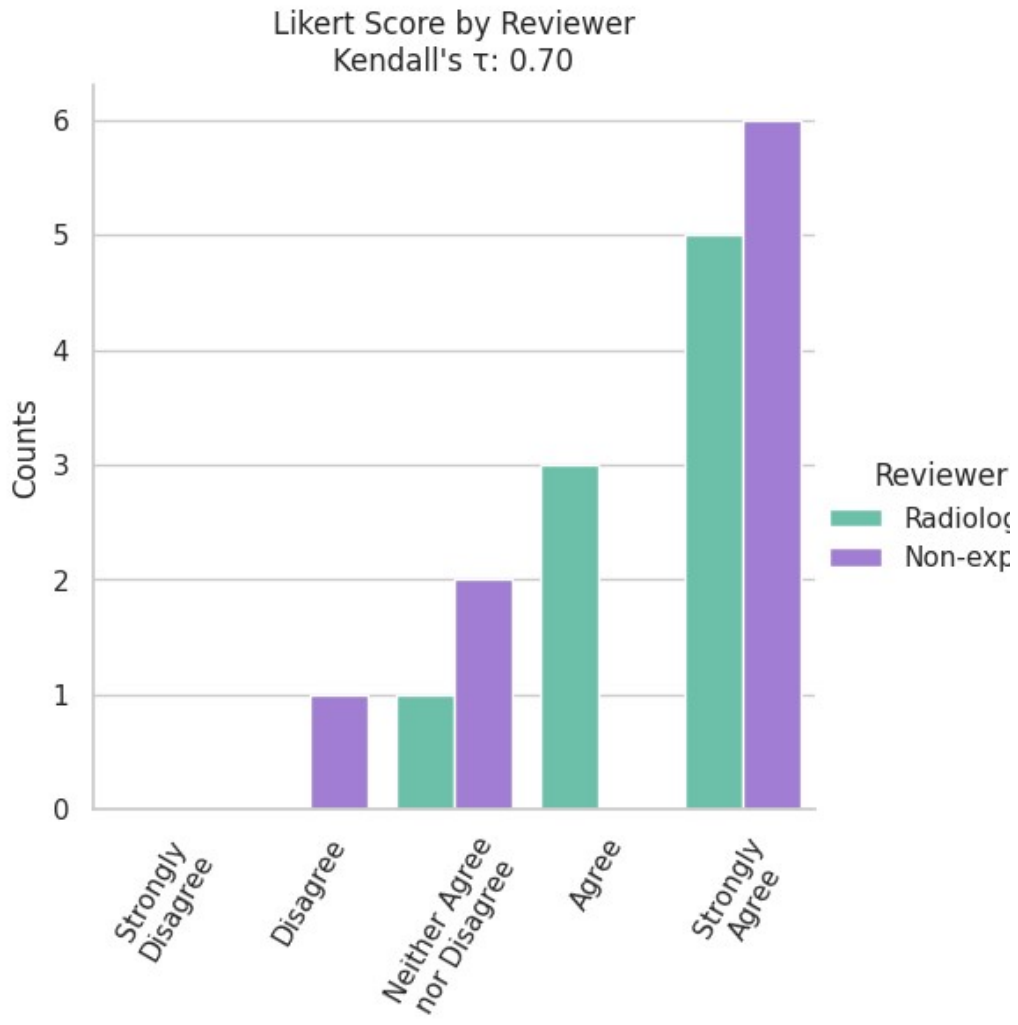

*Figure 18: Counts of Likert Scores for validation set of CT liver model by reviewers.*

## CT Liver Annotation

The radiologist and non-expert had both rated and corrected the liver segmentation in the 9 scans of the validations set. Figure 18 shows the reviewers had high correlation of their Likert scores. This task was similar to the liver and tumor segmentation task in the Medical Segmentation Decathlon. The nnU-Net[20] architecture won that challenge. The same metrics were used to compare our model to the state of the art performance, DSC and NSD where tolerance for NSD was set to 7mm. Table 9 shows the our model's performance on the validation in the TCGA-LIHC [18] collection, as well as the baseline state of the art performance of nnU-Net[20] on the MSD liver segmentation task dataset (https://doi.org/10.5281/zenodo.13851376).

## Limitations

One of the limitations of this study is the post-processing approach, which, while effective in removing small, isolated regions, does not eliminate all false positives. Some of these false positives may have been overlooked by radiologists, particularly when focusing on targeted regions, such as the kidneys. The dataset was generated through a fully automated process, with expert-corrected segmentations available for only a portion of the results. Due to limited resources, only 10% of the data underwent manual review and correction by a radiologist. This subset, while beneficial for improving segmentation accuracy, introduces a bias toward the AI model used to produce the masks. Additionally, the potential for human error persists even in the expert-corrected segmentations, and this remains a limitation in the dataset's overall accuracy and consistency.

*Table 9: Comparing nnUNet models performance with MSD Liver CT dataset. Our liver segmentation model was run on the TCGA-LIHC collection and compared with the 'ground truth' manually corrected annotations by Radiologist and Non-expert, where it also scored highly on MSD and NDS metrics. Our model is also evaluated on open source MSD-Liver CT dataset. Metrics are in mean, standard deviation provided for our model.*

| **Metric** | **Expert1** | **Non-Expert** | **MSD Liver-CT** |
|---|---|---|---|
| ***DSC*** | 0.97 (±0.07) | 0.97 (±0.08) | *0.94*(±0.04) |
| **NSD** | 0.96 (±0.09) | 0.95 (±0.11) | 0.95(±0.05) |
| **95% HD (mm)** | 5.66(±10.75) | 5.57(±12.71) | 18.01(±31.16) |

## Code Availability

The AI model weights are available on zenodo.org. Jupyter notebook code to reproduce the analysis can be found on zenodo.org. The URLs are given in Table 10.

*Table 10: URLs for model weights and analysis code*

| Task | Model Weights | Notebook Code |
|---|---|---|
| FDG PET/CT Lung and Lung Tumor Annotation | https://doi.org/10.5281/zenodo.8290054[28] | https://doi.org/10.5281/zenodo.13851020 |
| CT Lung Nodule Annotation | https://doi.org/10.5281/zenodo.8290146[33]<br>https://doi.org/10.5281/zenodo.8290168[32] | https://doi.org/10.5281/zenodo.13851336 |
| FDG PET/CT Breast Tumor Annotation | https://doi.org/10.5281/zenodo.8290054[28] | https://doi.org/10.5281/zenodo.13851345 |
| CT Kidneys, Tumors, and Cysts Annotation | https://doi.org/10.5281/zenodo.8277845[40] | https://doi.org/10.5281/zenodo.13851351 |
| MRI Prostate Annotation | https://doi.org/10.5281/zenodo.8290092[50] | https://doi.org/10.5281/zenodo.13851368 |
| MRI Liver Annotation | https://doi.org/10.5281/zenodo.8290123[60] | https://doi.org/10.5281/zenodo.13851371 |
| CT Liver Annotation | https://doi.org/10.5281/zenodo.8270230[63] | https://doi.org/10.5281/zenodo.13851376 |

## Acknowledgements

This project has been funded in whole or in part with Federal funds from the National Cancer Institute, National Institutes of Health, under Contract No. 75N91019D00024, Task Order No. 75N91020F0003. The content of this publication does not necessarily reflect the views or policies of the Department of Health and Human Services, nor does mention of trade names, commercial products or organizations imply endorsement by the U.S. Government.

The results published here are in whole or part based upon data generated by the TCGA Research Network: http://cancergenome.nih.gov/

Sharing of the QIN-Breast [13,14] collection is funded in part or whole by the NCI grand U01 CA142565

Sharing of the QIN-Prostate-Repeatability collection is funded in part or whole by the NCI grand U01 CA151261

## Author contributions

Gowtham Murugesan was responsible for development of the lung, breast, and kidney AI models. Gowtham also performed analysis of the AI vs QA annotations and drafted the manuscript. Diana McCrumb was responsible for the development of the prostate AI model and was the primary editor of the manuscript. Rahul Soni was responsible for development of the Liver AI models. Mariam Aboian reviewed and corrected the AI annotations as a radiologist. Tej Verma reviewed and corrected the AI annotations as a non-expert. Fatima Memon reviewed and corrected the AI annotations as a radiologist. David Clunie ensured correct DICOM-SEG encoding. Andrey Fedorov managed dataset compatibility and integration with IDC collections. Andrey also reviewed the manuscript. Linmin Pei managed the project for the NCI and reviewed the manuscript. Keyvan Farahani, Ulrike Wagner, and Stephen Moore reviewed and provided guidance for the project. Jeff Van Oss served as PI for the project and was responsible for converting files to and from DICOM, managing deliverables of code and annotations, and editing the manuscript.

## Competing interests

There are no conflicts of interest.